\documentclass[aps,prl,twocolumn,superscriptaddress,showpacs,10pt,floatfix]{revtex4}
\usepackage{graphicx} 
\begin{document}

\newcommand{\lnee}{\emph{$K^+\to l^+\nu e^+e^-$}}
\newcommand{\mnee}{\emph{$K^+\to \mu^+\nu e^+e^-$}}
\newcommand{\enee}{\emph{$K^+\to e^+\nu e^+e^-$}}
\newcommand{\eee}{\emph{$K_{e2ee}$}}
\newcommand{\mee}{\emph{$K_{\mu2ee}$}}
\newcommand{\lee}{\emph{$K_{l2ee}$}} \newcommand{\pee}{\emph{$K_{\pi
ee}$}} \newcommand{\lng}{\emph{$K\to l\nu\gamma$}}
\newcommand{\eng}{\emph{$K\to e\nu\gamma$}}
\newcommand{\mng}{\emph{$K\to\mu\nu\gamma$}}
\newcommand{\enp}{\emph{$K_{e3}$}\ }
\newcommand{\mnp}{\emph{$K_{\mu3}$}\ }
\newcommand{\lnp}{\emph{$K_{l3}$}\ } \newcommand{\rK}{\emph{$\langle
r_K^2\rangle$}} \newcommand{\LL}{{\cal L}}

\title{
{\bf \boldmath 
Experimental Study of the Radiative Decays
\mnee\ and \enee}
} 
\date{April 5, 2002}

\affiliation{Brookhaven National Laboratory, Upton, NY 11973, USA }
\affiliation{Department of Physics and Astronomy, University of New
Mexico, Albuquerque, NM 87131, USA} \affiliation{Department of Physics
and Astronomy, University of Pittsburgh, Pittsburgh, PA 15260, USA}
\affiliation{Institute for Nuclear Research of Russian Academy of
Sciences, Moscow 117 312, Russia} \affiliation{Paul Scherrer Institut,
CH-5232 Villigen, Switzerland} \affiliation{Physics Department, Yale
University, New Haven, CT 06511, USA} \affiliation{Physik-Institut,
Universit\"at Z\"urich, CH-8057 Z\"urich, Switzerland}

\author{A.~A.~Poblaguev} \affiliation{Institute for Nuclear Research
of Russian Academy of Sciences, Moscow 117 312, Russia}

\author{R.~Appel} \affiliation {Physics Department, Yale University,
New Haven, CT 06511, USA} \affiliation{Department of Physics and
Astronomy, University of Pittsburgh, Pittsburgh, PA 15260, USA}

\author{G.~S.~Atoyan} \affiliation{Institute for Nuclear Research of
Russian Academy of Sciences, Moscow 117 312, Russia}

\author{B.~Bassalleck} \affiliation{Department of Physics and
Astronomy, University of New Mexico, Albuquerque, NM 87131, USA}

\author{D.~R.~Bergman} \altaffiliation[{\scalebox{0.9}[1.0]{Present
address:}}]{Rutgers University, Piscataway, NJ 08855.}  \affiliation
{Physics Department, Yale University, New Haven, CT 06511, USA}

\author{N.~Cheung} \affiliation{Department of Physics and Astronomy,
University of Pittsburgh, Pittsburgh, PA 15260, USA}

\author{S.~Dhawan} \affiliation {Physics Department, Yale University,
New Haven, CT 06511, USA}

\author{H.~Do} \affiliation {Physics Department, Yale University, New
Haven, CT 06511, USA}

\author{J.~Egger} \affiliation{Paul Scherrer Institut, CH-5232
Villigen, Switzerland}

\author{S.~Eilerts} \altaffiliation[{\scalebox{0.9}[1.]{Present
address:}}]{The Prediction Co., Santa Fe, NM 87501.}
\affiliation{Department of Physics and Astronomy, University of New
Mexico, Albuquerque, NM 87131, USA}

\author{W.~Herold} \affiliation{Paul Scherrer Institut, CH-5232
Villigen, Switzerland}

\author{V.~V.~Issakov} \affiliation{Institute for Nuclear Research of
Russian Academy of Sciences, Moscow 117 312, Russia}

\author{H.~Kaspar} \affiliation{Paul Scherrer Institut, CH-5232
Villigen, Switzerland}

\author{D.~E.~Kraus} \affiliation{Department of Physics and Astronomy,
University of Pittsburgh, Pittsburgh, PA 15260, USA}

\author{D.~M.~Lazarus} \affiliation {Brookhaven National Laboratory,
Upton, NY 11973, USA }

\author{P.~Lichard} \affiliation{Department of Physics and Astronomy,
University of Pittsburgh, Pittsburgh, PA 15260, USA}

\author{J.~Lowe} \affiliation{Department of Physics and Astronomy,
University of New Mexico, Albuquerque, NM 87131, USA}

\author{J.~Lozano} \altaffiliation[{\scalebox{0.9}[1.]{Present
address:}}]{University of Connecticut, Storrs, CT 06269.}
\affiliation {Physics Department, Yale University, New Haven, CT
06511, USA}

\author{H.~Ma} \affiliation {Brookhaven National Laboratory, Upton, NY
11973, USA }

\author{W.~Majid} \altaffiliation[{\scalebox{0.9}[1.]{Present
address:}}]{LIGO/Caltech, Pasadena, CA 91125.}  \affiliation {Physics
Department, Yale University, New Haven, CT 06511, USA}

\author{S.~Pislak} \altaffiliation[{\scalebox{0.9}[1.]{Present
address:}}]{Phonak AG, CH-8712 St\"afa, Switzerland.}
\affiliation{Physik-Institut, Universit\"at Z\"urich, CH-8057
Z\"urich, Switzerland} \affiliation {Physics Department, Yale
University, New Haven, CT 06511, USA}

\author{P.~Rehak} \affiliation {Brookhaven National Laboratory, Upton,
NY 11973, USA }

\author{A.~Sher} \affiliation{Department of Physics and Astronomy,
University of Pittsburgh, Pittsburgh, PA 15260, USA}

\author{J.~A.~Thompson} \affiliation{Department of Physics and
Astronomy, University of Pittsburgh, Pittsburgh, PA 15260, USA}

\author{P.~Tru\"ol} \affiliation{Physik-Institut, Universit\"at
Z\"urich, CH-8057 Z\"urich, Switzerland} \affiliation {Physics
Department, Yale University, New Haven, CT 06511, USA}

\author{M.~E.~Zeller} \affiliation {Physics Department, Yale
University, New Haven, CT 06511, USA}

\begin{abstract}
Experiment 865 at the Brookhaven AGS obtained 410 \enee\ and 2679
\mnee\ events including 10\% and 19\% background. The branching ratios
were measured to be
$(2.48\pm0.14\textit{(stat.)}\pm0.14\textit{(syst.)})\times10^{-8}$
($m_{ee}>150~\mathrm{MeV}$) and $(7.06\pm0.16\pm0.26)\times10^{-8}$
($m_{ee}>145~\mathrm{MeV}$), respectively. Results for the decay form
factors are presented.
\end{abstract}


\maketitle

Chiral Perturbation Theory (ChPT) \cite{ChPT} has been a successful
approach to describing the decays of pseudoscalar mesons.  In the ChPT
program radiative kaon decays can serve both as an important test and
a source of input parameters for the theory.  While the decay modes
$K^+\rightarrow e^+ \nu \gamma$ $(K_{e2\gamma})$ and $\mu^+ \nu
\gamma$ $(K_{\mu2\gamma})$ have allowed some study of the form factors
involved \cite{Keng1,Keng2,Kmng}, the decays $K^+\rightarrow l^+ \nu
e^+ e^-\ (K_{e2ee},\ K_{\mu2ee})$ allow a more detailed investigation
into the structure of these decays.  We report here on such an
investigation from Experiment 865 at the Brookhaven National
Laboratory AGS with a 100-fold increase in the number of events in the
former mode and 150-fold increase in the latter \cite{Klnee}.

\begin{figure}[b]
\includegraphics[width=8.5cm]{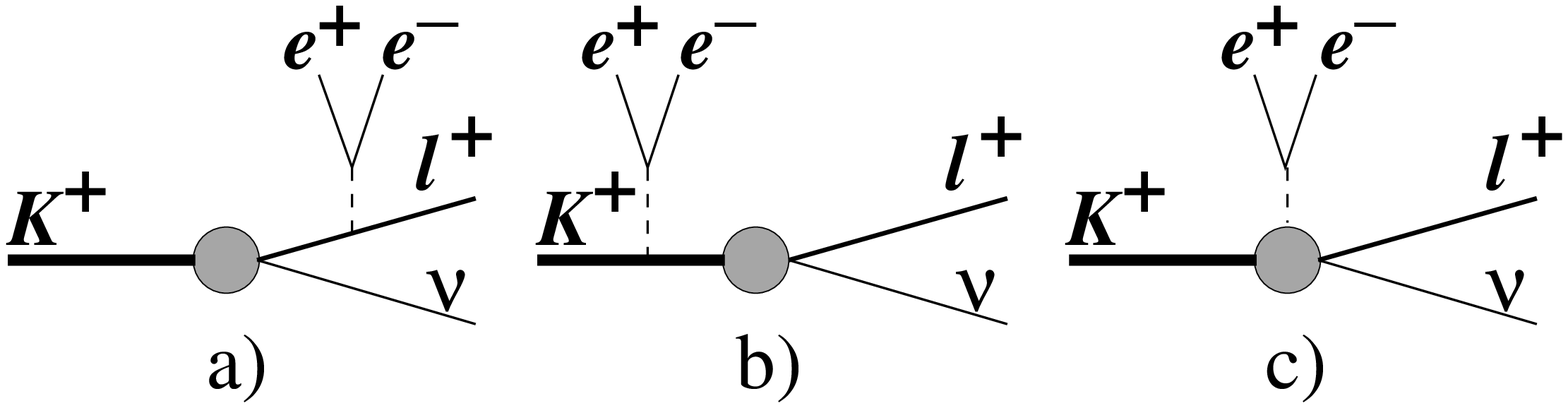}
\caption{\label{fig:graph} \lnee\ decay diagrams}
\end{figure}

The \eee\ and \mee\ decays are assumed to proceed via exchange of a
$W^+$-boson ($l^+\nu$) and photon ($e^+e^-$). The decay amplitude
\cite{Bardin,Bijnens} includes inner bremsstrahlung (IB) corresponding
to the tree diagrams in Fig. \ref{fig:graph}a,b, and structure
dependent (SD) radiation (Fig. \ref{fig:graph}c) parameterized by
vector $F_V$, axial $F_A$ and $R$ form factors. $K_{e2\gamma}$ and
$K_{\mu2\gamma}$ experiments were actually sensitive only to
$|F_V+F_A|$. $R$, which contributes only to decays with an
$e^+e^-$-pair, has not yet been measured.

	Inner bremsstrahlung is unambiguously predicted by the $K\to
l\nu$ amplitude and is proportional to the kaon decay constant
$F_K=160~\mathrm{MeV}$. We included the kaon electromagnetic form
factor in Fig. \ref{fig:graph}b (amplitude $A_4$ of
Ref. \cite{Bijnens}) in our definition of the IB term.  The IB
amplitude is negligible in \eee\ decay due to electron helicity
suppression, but dominates in \mee. It contributes about 60\% of the
total \mee\ branching ratio for invariant masses
$m_{ee}>145~\mathrm{MeV}$.  An additional 20\% comes from the IB and
SD amplitude interference%
, which makes it possible to measure the
signs of all form factors relative to $F_K$%
.

Generally form factors depend on $W^2$ and $q^2$, where $W$ and $q$
are 4-momenta of the $l^+\nu$-pair and of the photon ($e^+e^-$ pair),
respectively. In our analysis we assume the dominance of low lying
resonances \cite{Bardin}:
\begin{equation}
F_\textit{V,A,R}^{(q^2,W^2)} =
  F_\textit{V,A,R}\,/\,[(1-q^2/m_\rho^2)(1-W^2/\tilde{m}^2)]
\label{slopes}
\end{equation}
where $m_\rho=770~\mathrm{MeV}$, and
$\tilde{m}=m_{K^*}=892~\mathrm{MeV}$ for $F_V$ and
$\tilde{m}=m_{K_{1}}=1270~\mathrm{MeV}$ for $F_A$, $R$.  Only the
constants $F_V$, $F_A$, $R$, which we define in accordance with the
Particle Data Group (PDG) \cite{PDG}, will be the subject of our
analysis. The estimated uncertainties in the slope of form factors are
taken as the model errors.

We also included in the analysis a hypothetical tensor
amplitude
\begin{equation}
\frac{ieG_FV_{us}}{\sqrt{2}}F_T\epsilon^\mu q^\rho\,
\overline{u}_\nu(1+\gamma^5)\sigma^{\mu\rho}v_l
\label{tensor}
\end{equation}
 since a possible tensor interaction has been discussed in regard to
the $\pi\to e\nu\gamma$ \cite{ISTRA} and $K\to e\nu\pi^0$ \cite{Ke3}
experiments.

The experimental apparatus was constructed to search for the decay
$K^+\to\pi^+ \mu^+ e^-$ in flight from an unseparated 6 GeV/c $K^+$
beam, and has been described elsewhere \cite{E865:NIM}.  The
$K_{e2ee}$ and $K_{\mu2ee}$ data were obtained in a 1996 run
simultaneously with that for a measurement of $K^+\rightarrow \pi^+
e^+ e^-$\ $(K_{\pi ee})$ \cite{E865:piee}.  The trigger for these
modes allowed us to preselect events with three charged tracks,
including an $e^+e^-$ pair with high invariant mass
$m_{ee}$. Prescaled decays $K\to\pi\pi_D^0\ (K_{\pi2}),\
\mu\nu\pi_D^0\ (K_{\mu3}),\ e\nu\pi_D^0\ (K_{e3}),\ \pi\pi^0\pi_D^0\
(K_{\pi3})$ followed by Dalitz decay $\pi_D^0\to e^+e^-\gamma$ with
low $m_{ee}$ were used for normalization.

\begin{figure}
\scalebox{0.47}{%
\makebox[0pt][l]{\hspace{67pt}\raisebox{220pt}[0pt][0pt]%
{\LARGE \boldmath $K\to\mu\nu ee$}}%
\makebox[0pt][l]{\hspace{322pt}\raisebox{220pt}[0pt][0pt]%
{\LARGE \boldmath $K\to e\nu ee$}}%
\includegraphics{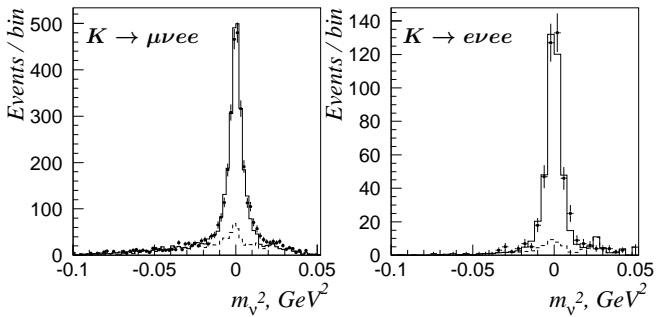}
}
\caption{\label{fig:mnu2} Missing mass distributions for \mee\ and
\eee\ decays.  Error bars are data, dashed lines indicate background,
and solid lines represent simulated distributions.}
\end{figure}

Off-line the $K_{e2ee}$ and $K_{\mu2ee}$ candidates were required to
have small missing neutrino mass, $m_\nu$.  This missing mass was
calculated using the measured decay product momenta and identities,
and the centroid of the incident kaon beam momentum determined from
$K^+\to\pi^+\pi^+\pi^-$ decays.  After correction for the decay vertex
dependence of the momentum, the kaon momentum resolution was
$\sigma_p/p=1.3\%$ and the angular resolution $\sigma_{\theta
x}=\sigma_{\theta y}=4~\mathrm{mrad}$.  The $m_\nu^2$ distribution for each decay modes is
displayed in Fig. \ref{fig:mnu2}.  A cut
$|m_\nu^2|<0.016~\mathrm{GeV}^2$ isolated \lee\ decays.

Background in both cases was dominated by accidental overlap tracks.
 Events with one of the tracks out of time gave model independent
 samples of the accidental background.  The samples were normalized by
 counting events with $m_\nu^2<-0.03~\mathrm{GeV}^2$ with
 normalization uncertainties 8\% for \mee\ and 25\% for \eee.  Other
 backgrounds and processes of interest were simulated using a GEANT3
 \cite{GEANT} based Monte Carlo.

The $K\to\pi ee$ decay is a potential background both for \eee\ and
\mee\ decays. To suppress it , each event was tested as \pee.  Events
giving effective ${\pi ee}$ mass and total momentum close to the beam
kaon values were removed.

Cuts on the invariant $e^+e^-$ mass $>145~\mathrm{MeV}$ (\mee) and
$>150~\mathrm{MeV}$ (\eee) removed backgrounds associated with large
branching ratio processes including a low mass $e^+e^-$-pair, {\em
e.g. $K\to\pi\pi_D^0$}.

After the $m_{ee}>150~\mathrm{MeV}$ cut 410 detected \eee\ candidate
events remain, including an estimated 35 accidental and 5 \pee\
background events.  The normalization sample was 86k \enp\ events and
2.3k $K_{\pi2}$ events.  The accuracy of the normalization was
estimated to be 4\%, including the 1.5\% error in the \enp\ branching
ratio, a 1\% trigger efficiency error, 2\% for the radiative
correction, and 3\% for reconstruction efficiency.  In evaluating the
normalization factor we assumed that the $\pi^0\to ee\gamma$ branching
ratio is 1.184\%, from the QED calculation \cite{Lautrup:1971ew} with
a form factor slope $0.032\pm0.004$ \cite{PDG}.

The scintillation hodoscope embedded in the muon stack behind 40 cm of
 iron \cite{E865:NIM} was used for additional $\mu/\pi$ separation
 in \mee\ analysis.  The $\mu/\pi$ detection efficiencies in the
 hodoscope were studied using $K_{\pi2}$ and $K_{\mu3}$ events where
 the $\pi^0_D$ was fully reconstructed, {\it i.e.}, the resulting
 photon was observed and included in the kinematic reconstruction.
 The average efficiency of detecting muons with momentum greater than
 0.9 GeV/c was 90\%, while the probability of the misidentifying the
 pion as a muon was about 20\%.  Because GEANT simulation predicts a
 larger pion misidentification, the correction factor 0.85 was
 applied to the simulated efficiency of the pion detection.  To reduce
 the uncertainty of this correction, the $K_{\pi2}$ decays in the
 normalization sample were additionally suppressed by a cut on
 $E_{\pi}<200~\mathrm{MeV}$ (calculated in the kaon center of
 mass). The final normalization sample contained 20.5k events
 including 16.2k \mnp, 3k $K_{\pi3}$, 0.8k $K_{\pi2}$, and 0.4k
 accidentals.  The normalization accuracy was estimated to be 5\%
 including 2.5\% from the \mnp\ branching ratio, 1\% from the trigger
 efficiency, 2\% from radiative corrections, 3\% from reconstruction
 efficiency and 2\% from efficiency of $\mu/\pi$ separation.  The
 total number of the selected \mee\ events with
 $m_{ee}>145~\mathrm{MeV}$ was 2679, including an estimated background
 of 355 accidentals, 126 \pee, and 33 $K\to\pi\pi_D^0\pi_D^0$ events.

\begin{table*}
\caption{\label{tab:results} Fit to E865 data. The errors are
$\pm\textit{stat.}\pm\textit{syst.}\pm\textit{model}$.  Results for
the form factors $F_V$,$F_A$, and $R$ (in units of $10^{-3}$) are
given for the fixed $F_K=160~\mathrm{MeV}$ and $F_T=0$. Because the
measurements of the form factors are correlated their linear
combinations are also presented.  Branching ratios BR (in units of
$10^{-10}$) were actually measured for $m_{ee}>145~\mathrm{MeV}$
(\mee) and $m_{ee}>150~\mathrm{MeV}$ (\eee).  All other values of BR
are extrapolations.  }
\begin{ruledtabular}
\begin{tabular}{l|c@{$\pm$}c@{$\pm$}c@{$\pm$}c|
                  c@{$\pm$}c@{$\pm$}c@{$\pm$}c|
                  c@{$\pm$}c@{$\pm$}c@{$\pm$}c|| c} &
                  \multicolumn{4}{c|}{$K^+\to\mu^+\nu e^+e^-$} &
                  \multicolumn{4}{c|}{$K^+\to e^+\nu e^+e^-$} &
                  \multicolumn{4}{c||}{Combined Fit} & Expected \\
                  \hline \hline $F_V$ & 124 & 19 & 13 & ~4 & ~87 & 30
                  & ~8 & ~5 & 112 & 15 & 10 & ~3 & { 96
                  \footnotemark[1]} \\ $F_A$ & ~31 & 21 & 14 & ~5 &
                  ~38 & 29 & 11 & ~3 & ~35 & 14 & 13 & ~3 &
                  { $41\pm~6$ \footnotemark[2] } \\ $R$ & 235 & 25 &
                  14 & 12 & 227 & 20 & 10 & ~8 & 227 & 13 & 10 & ~9 &
                  { $230\pm34$ \footnotemark[3]} \\ \hline $F_V+F_A$ &
                  155 & 25 & 21 & ~5 & 125 & 38 & 12 & ~3 & 147 & 21 &
                  15 & ~4 & { $\pm|144\pm9|$ \footnotemark[4]} \\
                  $F_V-F_A$ & ~93 & 32 & 17 & ~7 & ~50 & 44 & 15 & ~7
                  & ~77 & 20 & 19 & ~6 & { $102\pm74$
                  \footnotemark[5]} \\ $R+F_V$ & 359 & 36 & 20 & 14 &
                  314 & 34 & 11 & 12 & 338 & 19 & 15 & 11 & \\ $R-F_V$
                  & 111 & 26 & 18 & 11 & 139 & 37 & 12 & ~5 & 114 & 20
                  & 14 & ~8 & \\ $R+F_A$ & 265 & ~9 & 14 & ~7 & 265 &
                  14 & 10 & ~6 & 262 & ~6 & ~9 & ~6 & \\ $R-F_A$ & 204
                  & 46 & 25 & 17 & 189 & 48 & 18 & 10 & 191 & 27 & 22
                  & 12 & \\ \hline SL & \multicolumn{4}{c| }{ $11\%$ }
                  & \multicolumn{4}{c| }{ $36\%$ } &
                  \multicolumn{4}{c||}{ $12\%$ } & \\ \hline \hline
                  $F_K$ (MeV) & 157~& 7 & 5 & 0.3~ &
                  \multicolumn{4}{c|}{---} & 157~& 5 & 4 & 0.2~ & 160
                  \\ $F_T$ &$~-6~$ & 13 & ~8 & ~1 &$( 32^2$ & $37^2$ &
                  $25^2$ & $ 6^2)^{1/2}$ &$~-4~$ & 7 & 7 & 0.4~ & 0 \\
                  \hline \hline $BR_\textrm{total}$ &
                  \multicolumn{4}{c|}{---} &
                  \multicolumn{2}{c@{$\pm$}}{($1730^{+630}_{-540}$} &
                  90~& 80) \footnotemark[7] &
                  \multicolumn{4}{c||}{---} & \\ $BR_{m_{ee}>0.140}$
                  &(793 &~18 & 28 & 0.5) \footnotemark[7] &(291 & 16 &
                  17 & 0.7) \footnotemark[7] &
                  \multicolumn{4}{c||}{---} & {$1300\pm400
                  ~~/~~300^{+300}_{-150}$ \footnotemark[6]} \\
                  $BR_{m_{ee}>0.145}$ &\phantom{(}706 &~16 & 26 &
                  0.4\phantom{) \footnotemark[7]} &(270 & 15 & 16 &
                  0.4) \footnotemark[7] & \multicolumn{4}{c||}{---} &
                  \\ $BR_{m_{ee}>0.150}$ &(628 &~14 & 23 & 0.3)
                  \footnotemark[7] &\phantom{(}248 & 14 & 14 &
                  0.2\phantom{) \footnotemark[7]} &
                  \multicolumn{4}{c||}{---} & \\
\end{tabular}
\end{ruledtabular}
\begin{flushleft}
\footnotemark[1]{~Theoretical value (axial anomaly)
                    $F_V/M_K=\sqrt{2}/8\pi^2 F$ \cite{Bardin,Bijnens}.}\\
\footnotemark[2]{~ChPT to ${\cal O}(p^4)$
                    extrapolation from $\pi\to e\nu\gamma$ \cite{Bijnens,PDG}.
                    }\\
\footnotemark[3]{~$R/M_K=(1/3)F_K\langle
                    r_K^2\rangle$ \cite{Bardin}, with the experimental
                    value $\langle
                    r_K^2\rangle=0.34\pm0.05~\mathrm{fm}^2$
                    \cite{rK}.}\\ 
\footnotemark[4]{~$K\to e\nu\gamma$
                    and $K \to\mu\nu\gamma$ experimental data
                    \cite{Keng1,Keng2,Kmng}, corrected with the slopes
                    (\ref{slopes}) of the form factors
                    \cite{Comment}.} \\
\footnotemark[5]{~$K\to\mu\nu\gamma$ experimental
                    data \cite{Kmng}.} \\ 
\footnotemark[6]{~Previous
                    \mnee\ / \enee\ experimental data \cite{Klnee}.}  \\ 
\footnotemark[7]{~Extrapolated value.}
\end{flushleft}
\end{table*}

In order to fit the data, we have used the logarithmic likelihood
function
\begin{eqnarray}
   \LL & = &\sum_i2\left[m_i-n_i+m_i\ln(n_i/m_i)\right] \nonumber \\
 &-&\frac{(N_{MC}-N_{MC}^{(0)})^2}{\sigma_{MC}^2}
 -\frac{(N_{A}-N_{A}^{(0)})^2}{\sigma_{A}^2}
\label{chi2}
\end{eqnarray}
Here, $m_i$ and $n_i(F_V,F_A,R,N_{MC},N_A)$ are the measured and
calculated numbers of events in $i$-th bin of the 5-dimensional phase space (247 \mee\ and 147 \eee\ bins). We explicitly included the
uncertainties, $\sigma_{MC}$ and $\sigma_A$, in the Monte Carlo and
accidental normalizations, $N_{MC}$ and $N_A$, respectively in the
fit.  $N_{MC}$ and $N_A$ were regarded as independent parameters,
while $N_{MC}^{(0)}$ and $N_A^{(0)}$ are the expected values from our
studies discussed above.

The consistency between data and simulation was evaluated by
significance level (SL) \cite{PDG}, {\em i.e} the probability that a
random repeat of the experiment would observe a smaller $\LL$,
assuming the model is correct. To calculate the SL we simulated the
expected distribution of $\LL$. In this calculation we explicitly
accounted for possible variations of $n_i$ due to the finite Monte Carlo
and accidental statistics.

Fit results are summarized in Table \ref{tab:results}.  In the
combined fit the likelihood function used was the sum of \mee\ and
\eee\ likelihood functions (\ref{chi2}). Expectations for the measured
values, included in Table \ref{tab:results}, are based on the previous
$K_{l2\gamma}$ and \lee\ experiments, ChPT extrapolations of the
$\pi\to e\nu\gamma$ measurements, and theoretical predictions.

We considered three main contributions to the systematic error:
{\em(i)} uncertainty of the detector efficiencies; {\em(ii)}
uncertainty of the normalizations; {\em(iii)} errors due to the finite
statistics of the accidental and Monte Carlo samples.  All three
contributions are comparable, and their quadratic sums are shown in
Table \ref{tab:results}.

The model errors in Table \ref{tab:results} correspond to the
uncorrelated sum of the 30\% possible fluctuation of each slope of the
form factors given in Eq. (\ref{slopes}). Uncertainty in the $dR/dq^2$
dominates the model errors of all form factors.

We include in Table \ref{tab:results} both measured and extrapolated
branching ratios.  The \mee\ total branching ratio is expected to be
$2.5\times10^{-5}$ \cite{Bijnens}.  Because of our cut of
$m_{ee}>145~\mathrm{MeV}$, and because \mee\ has a large probability
of events at low $m_{ee}$ due to its being dominated by IB, we are
unable to determine the total branching ratio for that
mode.

To evaluate the sensitivity of our data to the IB term, we have made a
fit in which $F_K$ in the definition of IB amplitude was regarded as a
free parameter.  The resulting value of $F_K$ in Table \ref{tab:results},
being consistent with the expected value, also serves as 
a check of the normalization based on the \mnp\ decay.
The normalization based on the \enp\ decay can not be checked in the
same way in the \eee\ analysis. However, we can alternatively
determine that normalization by considering it as an unconstrained
parameter in the combined fit. The ratio of this normalization to that
obtained in \enp\ was found to be $(0.93\pm0.12)/(1.0\pm0.04)$.

Using the current algebra relationship \cite{Bardin} between form
factor $R$ and kaon charge radius $\langle r^2_K\rangle$,
$R = (1/3)M_KF_K\langle r^2_K\rangle$,
where $M_K$ is the kaon mass, we can calculate $\langle
r_K^2\rangle=0.333\pm0.027~\mathrm{fm}^2$ in agreement with the direct
measurement $0.34\pm0.05~\mathrm{fm}^2$ \cite{rK}. The difference
between our value and the experimental value of the pion charge radius
$\langle r_\pi^2\rangle=0.439\pm0.008~\mathrm{fm}^2$ \cite{rpi} does
not agree well with the ChPT prediction to ${\cal O}(p^4)$: $\langle
r_\pi^2\rangle - \langle
r_K^2\rangle=(1/32\pi^2F^2)\ln{(M_K^2/m_\pi^2)}= 0.036~\mathrm{fm}^2$
\cite{Bijnens}, where $F=92.4~\mathrm{MeV}$.

If we assume that the form factors are constant instead of varying
according to Eq. (\ref{slopes}) we obtain for the mean values:
$F_V=0.131$, $F_A=0.034$, $R=0.257$.  The value of $\LL$ indicates
that constant form factors are less likely by a factor of 3.6 than
those of Eq. (\ref{slopes}).

\begin{figure}
\includegraphics[width=5.7cm]{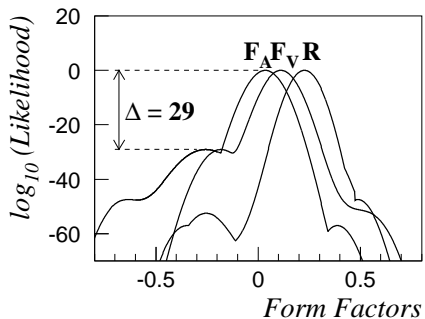}
\caption{\label{fig:ll} Likelihood functions for form factors $F_V$,
$F_A$, and $R$.  }
\end{figure}

While measurements of the \eee\ decay can only determine the signs of
the form factors relative to one another, due to the interference
between the IB and SD amplitudes, knowledge of \mee\ allows us to
establish the signs relative to $F_K$.  For the combined fit, the
discrimination against wrong sign combinations is illustrated in
Fig. \ref{fig:ll}.

The presentation of the tensor form factor for \eee\ in Table
\ref{tab:results} underlines that it is $F_T^2$ which is  measured in
this mode. Our analysis of the \mee\ decay is, however, sensitive to the sign of $F_T$. We did not find any evidence for the presence of
a tensor term Eq. (\ref{tensor}).  Our result does not rule out
the value $F_T=-0.0056\pm0.0017$ \cite{peng} invoked as a possible interpretation  \cite{AAP:tensor} of the $\pi\to
e\nu\gamma$ data from Ref. \cite{ISTRA}.  Assuming that the \enp\
tensor form factor is related to $F_T$ as $f_T/f_+=3.8F_T$
\cite{Ke3tensor}, we find $f_T/f_+=-0.02\pm0.04$ which strongly
disagrees with value $0.53^{+0.09}_{-0.10}\pm0.10$ of Ref. \cite{Ke3}.
A value compatible with 0 was also obtained in recent \lnp\
experiments \cite{Ke3:new}.

To summarize, we have measured the \eee\ and \mee\ branching ratios:
$(2.48\pm0.14\textit{(stat.)}\pm0.14\textit{(syst.)})\times10^{-8}$
($m_{ee}>150~\mathrm{MeV}$) and $(7.06\pm0.16\pm0.26)\times10^{-8}$
($m_{ee}>145~\mathrm{MeV}$), respectively. For the first time all
\lee\ form factors were unambiguously measured:
$F_V=0.112\pm0.015\pm0.010\pm0.003\textit{(model)}$,
$F_A=0.035\pm0.014\pm0.013\pm0.003$,
$R=0.227\pm0.013\pm0.010\pm0.009$.  Our analysis was especially
sensitive to the sum of axial form factors
$R+F_A=0.262\pm0.006\pm0.009\pm0.006$.  We did not find any
inconsistency between \eee\ and \mee\ form factors.  The measured
inner bremsstrahlung contribution $F_K=157\pm5\pm4~\mathrm{MeV}$
agrees well with the theoretical expectation. No evidence of the
tensor amplitude was found: $F_T=-0.004\pm0.007\pm0.007$.  Our study
of the form factors is more detailed but consistent with previous
$K(\pi)\to l\nu\gamma$ and \lnee\ experiments.

\begin{acknowledgments}
We thank J.~Bijnens for his FORTRAN code describing the \lnee\ matrix
element.  We gratefully acknowledge the contributions to the success
of this experiment by the staff and management of the AGS at the
Brookhaven National Laboratory, and the technical staffs of the
participating institutions.  This work was supported in part by the
U. S. Department of Energy, the National Science Foundations of the
USA, Russia and Switzerland, and the Research Corporation.
\end{acknowledgments}

\bibliography{e865lnee}

\begin{thebibliography}{24}
\bibitem[{Com()}]{Comment}
\bibinfo{note}{Our estimate based on the reanalysis of the published
  experimental data, gives the following corrections to the
  $|F_V+F_A|$:~~$-0.012$ \cite{Keng1}, $-0.012$ \cite{Keng2}, $-0.008$
  \cite{Kmng}. The last value may be compared with the authors' one $-0.010$,
  obtained in the real fit of the available data.}
\bibitem[{pen()}]{peng}
\bibinfo{note}{The tensor couplings in the pion and kaon decays are related
  here assuming the same Cabibbo angle for the tensor and V-A amplitudes
  \cite{Gabrielli}.}
\bibitem[{Ke3({\natexlab{a}})}]{Ke3tensor}
\bibinfo{note}{We derived the factor 3.8 by comparing theoretical calculations
  of the tensor form factors in \lee \cite{Belyaev,Gabrielli} and \lnp
  \cite{Chizhov}. We do not discuss the confidence of the cited calculations
  nor did we evaluate the accuracy of the factor 3.8}.
\end{thebibliography}

\end{document}